\documentclass[prl,floatfix,twocolumn,showpacs,amsmath,amssymb]{revtex4}
\usepackage{graphicx}
\usepackage{dcolumn}
\usepackage{bm}

\begin{document}
\title{Magneto-elastic effects and magnetization plateaus in two dimensional
systems} 

\author{Samuele Bissola,$^{1}$ Valeria Lante,$^{1}$ Alberto Parola,$^{1}$
and Federico Becca$^{2}$}
\affiliation{
$^{1}$ Dipartimento di Fisica e Matematica, Universit\`a dell'Insubria, 
Via Valleggio 11, I-22100 Como, Italy \\ 
$^{2}$ CNR-INFM Democritos National Simulation Center and International 
School for Advanced Studies (SISSA), via Beirut 2-4, I-34014 Trieste, Italy}

\date{\today}

\begin{abstract}
We show the importance of both strong frustration and spin-lattice coupling 
for the stabilization of magnetization plateaus in translationally 
invariant two-dimensional systems. We consider a frustrated spin-1/2
Heisenberg model coupled to adiabatic phonons under an external magnetic field.
At zero magnetization, simple structures with two or at most four spins per 
unit cell are stabilized, forming dimers or $2 \times 2$ plaquettes, 
respectively. A much richer scenario is found in the case of 
magnetization $m=1/2$, where larger unit cells are formed with non-trivial 
spin textures and an analogy with the corresponding classical Ising model 
is detectable. Specific predictions on lattice distortions and local spin
values can be directly measured by X-rays and Nuclear Magnetic Resonance 
experiments.

\end{abstract}


\maketitle

A magnetization plateau occurs when the magnetization remains constant 
over a range $\Delta h$ of applied magnetic fields.
The width of the plateau can be expressed in term of the excitation spectrum
$\Delta h = E(M+1)-2E(M)+E(M-1)$,
where $E(M)$ is the total energy at fixed magnetization $M$ (measured in
units of $g\mu_B$).
This, in turn, implies that the energy (per site)
as a function of the magnetization (per site) displays a cusp-like
singularity in the thermodynamic limit.
In general, plateaus are absent in classical models with magnetic
ground states whenever the
magnetization is not collinear with the field.~\cite{lhuillier1}
Instead, they identify particularly stable quantum phases characterized by a 
spin gap. 

Usually, gaps in the excitation spectrum directly reflect the structure 
of the primitive cell of the lattice according to the commensurability 
condition~\cite{affleck} 
\begin{equation}
\ell S(1-m)={\rm integer}
\label{oshi}
\end{equation}
where $S$ is the magnitude of the spin, $m$ the magnetization per site
in units of $g\mu_BS$ (i.e., the average on-site spin component parallel to 
the magnetic field), and $\ell$ the number of spins in the primitive cell.
According to Eq.~(\ref{oshi}), magnetization plateaus in models with a single 
spin-1/2 per unit cell may occur only as a consequence of the {\it spontaneous}
breaking of the translational symmetry, leading to an $\ell$-fold degenerate 
ground state.  Although this relation has been rigorously proved in 
one-dimensional (1D) systems~\cite{affleck}, it is believed to have a 
much wider validity.~\cite{oshikawa} In fact, several analytical and 
numerical calculations in chains and ladders have definitively confirmed the 
presence of various magnetization plateaus at the expected 
positions.~\cite{okunishi,cabra,mila}
Moreover, few important cases of magnetization plateaus in higher dimensions 
are also known. The most relevant example is given by the orthogonal dimer 
Heisenberg model, which closely represents the structure of copper planes in 
${\rm SrCu_2(BO_3)_2}$.~\cite{miyahara} In this case, the theoretical 
predictions have been confirmed by experimental evidence of magnetization 
plateaus at $m=0$, $1/8$, and $1/4$.~\cite{kageyama}
Some evidence for a $m=1/3$ plateau has been proposed for the triangular 
lattice,~\cite{nishimori,ono} while in the square lattice, the $m=0$ 
properties of the $J_1{-}J_2$ model are still debated. 
Indeed, although for 
$J_2/J_1 \sim 1/2$ the ground state is believed to be disordered, the 
existence of a finite triplet gap is much less clear,~\cite{caprio,wen} 
casting some doubt as to the possibility of having an $m=0$ plateau.

The interest in the $J_1{-}J_2$ model has grown due to the recent discovery 
of two materials well described by a two-dimensional (2D) quantum 
antiferromagnet, i.e., ${\rm Li_2VOSiO_4}$ and 
${\rm VOMoO_4}$.~\cite{carretta,carretta2} Although the experimental 
magnetization curve of these compounds has not yet been considered,
the magnetization properties of the $J_1{-}J_2$ 
model have been recently considered by using exact diagonalization 
calculations, leading to some evidence in favor of a magnetization plateau at 
$m=1/2$.~\cite{zhitomirsky} This outcome has 
been interpreted as a consequence of the emergence of a $2 \times 2$ super-cell.
According to this scenario, inside each cell the magnetic moments acquire a 
preferential orientation along the direction of the magnetic field, leading 
to a configuration with three up and one down spin.~\cite{zhitomirsky}
Unfortunately, the presence of such a half-magnetization plateau is limited 
to a very narrow region close to $J_2/J_1 \sim 1/2$, revealing the
difficulty of stabilizing such a state in the pure spin model without other 
degrees of freedom. 
In this respect, the spin-lattice coupling represents one of the
most relevant physical mechanisms to enhance the stability of phases breaking 
some of the lattice symmetries, making it easier to identify these states in 
numerical studies of small lattices. 
From general arguments, the 
super-exchange couplings are ultimately generated by the virtual hopping of 
electrons through neighboring sites and strongly depend upon lattice 
distortions.~\cite{harrison}
The role of the spin-lattice coupling in frustrated spin systems has been 
extensively considered in the absence of an external magnetic field for both 
1D~\cite{poilblanc1,becca1} and 2D systems.~\cite{becca2}
More recently, the importance of lattice distortions for stabilizing
magnetization plateaus has been discussed in a simple 1D spin 
model,~\cite{poilblanc2} and in a classical Heisenberg model on a pyrochlore 
lattice.~\cite{penc}

In this Letter we investigate the possible occurrence of magnetization 
plateaus in the frustrated spin-1/2 Heisenberg model on the square lattice
by the numerical analysis of the periodicity of the distortion pattern 
induced by spin-lattice coupling. The model is defined by
\begin{equation}
{\cal H} = 
\sum_{i,j} \left [ J(d_{ij}) {\bf S}_i \cdot {\bf S}_j + \frac{K(d_{ij}^0)}{2}
\left ( \frac{\|\delta {\bf r}_i-\delta {\bf r}_j \|}{d_{ij}^{0}} \right )^2 
\right ]
\label{hamilt}
\end{equation}
where ${\bf S}_i$ is the spin-$1/2$ operator at site $i$,
$\delta {\bf r}_i$ is the displacement of atom $i$, assumed to be in the
plane, and $d_{ij} \equiv \Vert \mathbf{r}_i -
\mathbf{r}_j \Vert$ is the distance between atoms $i$ and $j$. The sum
runs over nearest [i.e., $J_1=J(1)$], second [i.e., $J_2=J(\sqrt{2})$]
and third [i.e., $J_3=J(2)$] neighbor sites on a 2D square lattice, while 
$d_{ij}^{0}=\|{\bf R}_{i}^{0}-{\bf R}_{j}^{0}\|$ is the distance
between sites $i$ and $j$ in the undistorted lattice. Energy units 
are fixed by the choice $J_1=1$. 
In order to limit the number of parameters, only the elastic constant 
 between nearest-neighbor sites [$K=K(1)$] has been included. 
However, we checked 
that our conclusions are not qualitatively affected by considering 
further neighbor elastic constants. In transition metal compounds, the 
super-exchange theory combined with empirical dependences of hopping 
integrals on distance~\cite{harrison} leads to exchange integrals that vary 
like the inverse of the distance to a given power $\alpha_{\mu}$ ($\mu=1,2,3$). 
For small displacements we can write:
\begin{equation}\label{jj}
J(d_{ij})=J(d_{ij}^0)
\left ( \frac{d_{ij}^0}{d_{ij}} \right )^{\alpha_{\mu}} \simeq
J (d_{ij}^0) \left ( 1 - \alpha_{\mu} \frac{\delta d_{ij}}{d_{ij}^{0}} 
\right ),
\end{equation}
with $\delta d_{ij} = d_{ij} - d_{ij}^0 \sim ({\bf R}_{i}^{0}-{\bf R}_{j}^{0})
\cdot (\delta {\bf r}_{i}-\delta {\bf r}_{j})/d_{ij}^0$.
Since the Hamiltonian is invariant under the rescaling 
$\alpha_{\mu}=\lambda\alpha_{\mu}$, $K=\lambda ^2 K$ and 
$\delta {\bf r}_{i}=\delta {\bf r}_{i}/\lambda$, $K$ can be fixed.
To study this system, we adopt the Lanczos diagonalization technique
in finite clusters with periodic boundary conditions, which allows for an 
unbiased determination of the lowest-energy configuration at fixed 
distortion pattern. In order to determine the optimal configuration of lattice 
displacements $\delta {\bf r}_{i}$ we use an iterative procedure.~\cite{becca2}
Since the local displacement of each spin from its equilibrium
position is considered, we are able to describe all kinds of distortions,
including the dilation and the shrinking of the lattice.
We performed systematic calculations at different frustrations and
lattice couplings in a $4 \times 4$ cluster at 
$m=0$ and $m=1/2$, i.e., at the magnetization values where plateaus
are commonly expected in this geometry. 

At vanishing magnetization, the classical $J_1{-}J_2{-}J_3$ model on a rigid 
lattice shows four magnetically ordered states: N\'eel with momentum 
($\pi,\pi$), collinear with momenta ($\pi,0$) and ($0,\pi$) and two 
helicoidal phases with momenta ($q,q$) and ($q,\pi$), ($\pi,q$), where $q$ 
varies continuously with the parameters in the 
Hamiltonian.~\cite{moreo,chub,ferrer} Quantum fluctuations, enhanced by the 
competing antiferromagnetic couplings, can drive the system away from these 
semiclassical behaviors and stabilize unconventional quantum phases without 
long-range magnetic order, which may eventually display a magnetization 
plateau. 
At $J_3=0$, the magnetically disordered regime is widely believed to occur 
around the maximally frustrated point (i.e., $J_2=1/2$), while at $J_2=0$ a 
non-classical phase appears between the N\'eel and the spiral state ($q,q$) 
close to $J_3 \sim 1/2$. However, in both cases, the nature of the disordered 
phases is still controversial and several proposals appear in the 
literature:~\cite{lhuillier} Valence bond crystal (VBC) columnar 
states~\cite{read,leung} or spin liquids.~\cite{caprio,caprio1}
Very recently a singlet state with plaquette order has been claimed to be 
stabilized around the line $J_2+J_3 \simeq J_1/2$ with 
$J_3 > 0$.~\cite{poilblanc3}

\begin{figure}
\includegraphics[width=0.48\textwidth]{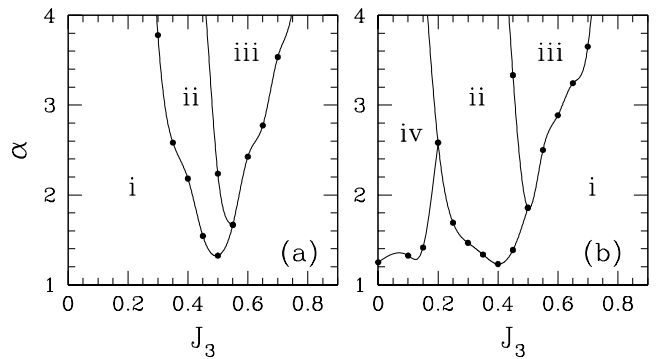}
\caption{\label{lattice1}
Ground-state phase diagram of the $J_1-J_2-J_3$ model for $m=0$, $K=10$,
and $J_2=0$ (a) or $J_2=0.5$ (b). See text for the precise description of the
various phases.}
\end{figure}

When the spin-lattice coupling is taken into account, we expect that each of 
the previous spatial orderings would lead to a characteristic distortion 
pattern which can be easily identified. In Fig.~\ref{lattice1} we present 
the results for the ground-state phase diagram obtained by Lanczos 
diagonalizations of the Hamiltonian~(\ref{hamilt}) in the $J_3$-$\alpha$ 
plane for $\alpha_{\mu}=\alpha$ and $K=10$. Two values of $J_2=0$ and $1/2$, 
representative of the weak and strong frustration regime, will be considered.

Among all possible lattice deformations, only four are stable at $m=0$, 
depending on the values of the parameters (see Fig.~\ref{lattice1}):
{\it{i}}) The square lattice with uniform bond lengths:
All spatial symmetries are preserved and the presence of the spin-phonon 
coupling $\alpha$ just leads to a renormalization of the bond lengths.
{\it{ii}}) The dimerized lattice with two different bonds in one direction 
and one bond in the other one. This phase breaks both $\pi/2$ rotation and
the translational symmetry along one direction and the ground state is 
four-fold degenerate. In this case, the commensurability condition~(\ref{oshi})
is satisfied leading to spin gap and a magnetization plateau.
{\it{iii}}) The square-plaquette phase, with dimerization 
in both directions. In this state, rotational symmetry is preserved since
a $2\times 2$ primitive cell is stabilized.
Also in this case the ground state is four-fold degenerate and the spin gap 
is finite. In any case, the plausible occurrence of spiral magnetic
ordering with incommensurate $q$ at large $J_3$ suggests to be cautious with 
the numerical results in small lattices for $J_3 \gtrsim 0.6$. 
For $J_2 = 1/2$, we also found {\it{iv}}) the rectangular phase with 
different bond lengths in the $x$ and $y$ directions. This phase breaks the 
$\pi/2$ rotational symmetry but is translationally invariant, as expected 
when collinear magnetic ordering is present. In this case the ground state 
is two-fold degenerate, leading to a finite-temperature phase 
transition.~\cite{weber}
Finally, the absence of distortions for low values of $J_3$ is consistent 
with a magnetic ground state displaying N\'eel order. 
In all of these cases, the local magnetization $\langle S_i^z \rangle$ 
is uniform in the whole lattice. 

In order to better understand the physical properties of the two intermediate 
phases, {\it{ii}}) and {\it{iii}}), we have analyzed the energy spectrum 
both in the $4\times 4$ and in the $6\times 6$ {\it rigid} lattice.  
Some evidence of translational symmetry breaking
emerges from the quasi degeneracy of the states with momenta $(0,0)$,
$(0,\pi)$ and $(\pi,0)$. However, the ordering of energy levels 
on these small systems is not able to discriminate between
a columnar dimer or a plaquette state, because both the $d$-wave, zero 
momentum state and the ($\pi,\pi$) $s$-wave singlet have comparable 
energies in these clusters. 

\begin{figure}
\includegraphics[width=0.48\textwidth]{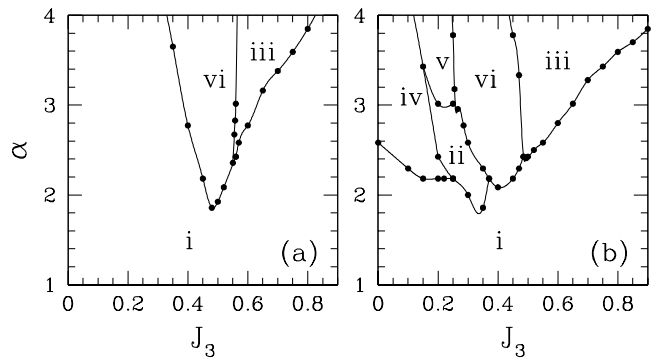}
\caption{\label{lattice2}
The same as in Fig.~\ref{lattice1} but for $m=1/2$.}
\end{figure}

Let us now move to the more interesting case of $m=1/2$. The phase diagram
is shown in Fig.~\ref{lattice2}. 
At small $J_3$, a semiclassical scenario is 
consistent with both the shape of the lattice and the observed uniform local
magnetization (i.e., $\langle S_i^z \rangle=1/4$). Indeed, an applied magnetic
field cants the spins which preserve a magnetic ordering in the plane
orthogonal to the field direction.
The numerical evaluation of the spin-spin correlation function
by means of Lanczos diagonalizations on the $4\times 4$ and $6\times 6$ 
rigid cluster is fully compatible with the semiclassical scenario both at 
small and large $J_2$, namely weak correlations along the magnetic field and 
marked peaks of the structure factor in the orthogonal plane. 
The analytical evaluation of the susceptibility via spin-wave theory
shows that, for weak phonon coupling, the lattice undergoes
global deformations but bond lengths and local magnetizations remain 
uniform throughout the lattice, thereby inhibiting magnetization plateaus.  

\begin{figure}
\includegraphics[width=0.4\textwidth]{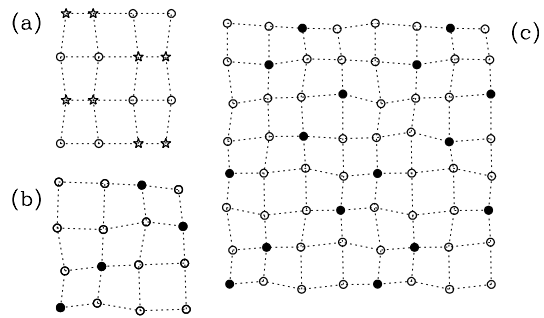}
\caption{\label{ising} 
Distortion pattern of the $J_1{-}J_2{-}J_3$ Heisenberg model on the 
$4 \times 4$ lattice at $m=1/2$ in the ``trapezoidal'' phase at $J_2=0.5$, 
$J_3=0.2$, and $\alpha=3.5$ (a) and in the ``classical Ising'' phase at 
$J_2=0.5$, $J_3=0.4$, and $\alpha=\sqrt{10}$ (b). Distortion pattern of the 
Ising model on the $8\times 8$ cluster at $J_2=0.5$, $J_3=0.4$, and
$\alpha=\sqrt{10}$ (c).
Stars mean $\langle S_i^z \rangle \sim 0$, while empty and full dots indicate
positive and negative local magnetization close or equal to the maximal 
value 1/2, respectively.}
\end{figure}

\begin{figure}
\includegraphics[width=0.42\textwidth]{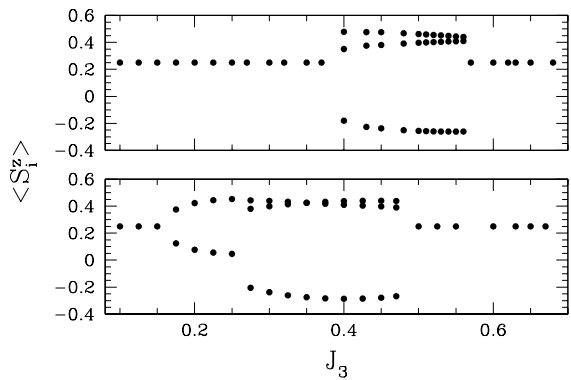}
\caption{\label{SZ}
Behavior of local magnetization $\langle S_i^z \rangle$ as a function of 
$J_3$ at $m=1/2$ and $\alpha=\sqrt{10}$ for $J_2=0$ (upper panel) 
and $J_2=0.5$ (lower panel).}
\end{figure}

By increasing the third neighbor coupling $J_3$, in addition to the same 
lattice deformations already found at zero magnetization, other two phases 
emerge (see Fig.\ref{lattice2}):
{\it{v}}) The ``trapezoidal'' phase characterized by a $4 \times 2$ primitive 
cell made of two aligned congruent isosceles trapezia, tilted by 
$\pi$ one respect to the other. In this case, the lattice dilates in one 
direction and exhibits three different bond lengths, still preserving a 
rectangular shape (see Fig.~\ref{ising}). The local magnetization is not 
uniform but it acquires two different values (see Fig.~\ref{SZ}). 
{\it{vi}}) The ``classical Ising'' phase characterized by a $4 \times 4$ 
unitary cell composed of four identical scalene trapezia which can be 
obtained one from the other by means of suitable rotations 
(see Fig.~\ref{ising}). The lattice is not squeezed in any direction, but 
develops a complex bond pattern with eight different bond lengths 
and three values of the local magnetization: Two positive and one negative,
with modulus close to the limiting value $S=1/2$ (see Fig.~\ref{SZ}). 
This rather rich texture of local magnetizations $\langle S_i^z \rangle$ is
stabilized for non-zero third neighbors coupling and can be directly tested 
by Nuclear Magnetic Resonance (NMR) experiments. Interestingly, this spin 
pattern has a direct interpretation on the basis of a purely classical Ising 
model. Indeed,
we checked that the complex topology previously described is exactly shared 
by a 2D Ising model with antiferromagnetic interactions up to third neighbors
and spin-lattice coupling at one-half of the saturation magnetization.
A numerical study of this model in the same parameter regime
has been carried out both in the $4 \times 4$ and in the $8 \times 8$ lattices
with similar results: The spin patterns is characterized by the requirement of
placing the down spins so to avoid first and third neighbors, while allowing 
a single second neighbor down spin, see Fig.~\ref{ising}.~\cite{note}
It should be noticed that this constraint  leads
to slightly different patterns in the $4 \times 4$ and the $8 \times 8$ 
clusters, because of periodic boundary conditions, and on the $4 \times 4$ is 
compatible with the results on the quantum Heisenberg model.
Moreover, also the distortion pattern of the classical model is extremely 
similar to that obtained in the quantum model (see Fig.~\ref{ising}).

It is worth noting that once the magnetic energy gain overcomes the elastic 
energy, the transition lines among the different phases depend only 
weakly on the spin-phonon coupling $\alpha$. This suggests that, in the 
thermodynamic limit, the distorted phases might remain stable up to
arbitrarily  
small $\alpha$, as can be analytically shown for the nearest-neighbor Ising 
model. In such a case, the $m=1/2$ magnetization plateau would be present, 
in the $J_1{-}J_2{-}J_3$ model, independently of the strength of spin-lattice
coupling.  

In summary, we showed 
that magnetization plateaus can be stabilized in the square lattice 
$J_1{-}J_2{-}J_3$ antiferromagnetic Heisenberg model in the presence of
spin-phonon coupling. Third neighbor coupling appears to be an essential 
ingredient for the appearance of plateaus both at zero and one-half 
magnetization. While at $m=0$ the lattice dimerizes as expected on the basis 
of a VBC phase, at $m=1/2$ a novel and complex distortion 
pattern characterizes the broken symmetry state. This phase can be faithfully 
interpreted in terms of a frustrated Ising model coupled to the lattice and 
appears as a genuine result not much affected by the finite size of the 
cluster we analyzed. The emerging scenario is considerably richer than
predicted in a seminal study of the $J_1-J_2$ model on a rigid 
lattice~\cite{zhitomirsky} and identifies a simple mechanism for the 
stabilization of complex spatial structures. The on-site magnetization 
displays a distinctive pattern which may be measured by NMR experiments in 
strongly frustrated 2D magnetic materials, like ${\rm Li_2VOSiO_4}$ or
${\rm VOMoO_4}$.

We acknowledge interesting discussion with P. Carretta and D. Poilblanc.
This work has been partially supported by COFIN 2005 and CNR-INFM.


\begin{thebibliography}{99}

\bibitem{lhuillier1} C. Lhuillier and G. Misguich, {\it High Magnetic Fields} 
   (Springer Lecture Notes in Physics, vol. 595) ed. C. Berthier, L.P. L\'evy, 
   and G. Martinez, p.161 (cond-mat/0109146).
\bibitem{affleck} M. Oshikawa, M. Yamanaka, and I. Affleck, \prl {\bf 78}, 1984
   (1997).
\bibitem{oshikawa} M. Oshikawa, \prl {\bf 84}, 1538 (2000).
\bibitem{okunishi} K. Okunishi and T. Tonegawa, J. Phys. Soc. Japan {\bf 72},
   479 (2003).
\bibitem{cabra} D.C. Cabra, A. Honecker, and P. Pujol, \prl {\bf 79}, 5126 
   (1997); Eur. Phys. J. B {\bf 13}, 55 (2000).
\bibitem{mila} A. Honecker, F. Mila, and M. Troyer, Eur. Phys. J. B {\bf 15}, 
   227 (2000).
\bibitem{miyahara} S. Miyahara and K. Ueda, \prl {\bf 82}, 3701 (1999).
\bibitem{kageyama} H. Kageyama {\it et al.}, \prl {\bf 82}, 3168 (1999).
\bibitem{nishimori} H. Nishimori and S. Miyashita, J. Phys. Soc. Japan {\bf 55},
   4448 (1986); A. Honecker, J. Phys.: Condens. Matter {\bf 11}, 4697 (1999).
\bibitem{ono} T. Ono {\it et al.}, \prb {\bf 67}, 104431 (2003).
\bibitem{caprio} L. Capriotti {\it et al.}, \prl {\bf 87}, 097201 (2001).
\bibitem{wen} Y. Ran and X.-G. Wen, cond-mat/0609620.
\bibitem{carretta} R. Melzi {\it et al.}, \prl {\bf 85}, 1318 (2000); \prb 
   {\bf 64}, 024409 (2001).
\bibitem{carretta2} P. Carretta {\it et al.}, \prb {\bf 66}, 094420 (2002).
\bibitem{zhitomirsky} M.E. Zhitomirsky, A. Honecker, and O.A. Petrenko, 
   \prl {\bf 85}, 3269 (2000).
\bibitem{harrison} W.A. Harrison, {\it Electronic structure and the properties
   of solids} (Dover, New York, 1980).
\bibitem{poilblanc1} J. Riera and D. Poilblanc, \prb {\bf 59}, 2667 (1999).
\bibitem{becca1} F. Becca, F. Mila, and D. Poilblanc, \prl {\bf 91}, 067202 
   (2003).
\bibitem{becca2} F. Becca and F. Mila, \prl {\bf 89}, 037204 (2002).
\bibitem{poilblanc2} T. Vekua {\it et al.}, \prl {\bf 96}, 117205 (2006).
\bibitem{penc} K. Penc, N. Shannon, and H. Shiba, \prl {\bf 93}, 197203 (2004).
\bibitem{moreo} A. Moreo, E. Dagotto, T. Jolicoeur, and J. Riera, \prb 
   {\bf 42}, 6283 (1990).
\bibitem{chub} A. Chubukov, \prb {\bf 44}, 392 (1991).
\bibitem{ferrer} J. Ferrer, \prb {\bf 47}, 8769 (1993).
\bibitem{lhuillier} See for instance, G. Misguich and C. Lhuillier in 
   {\it Frustrated spin systems}, H.T. Diep editor, World-Scientific (2005) 
   and Refs. therein.
\bibitem{read} N. Read and S. Sachdev, \prl {\bf 62}, 66 (1989).  
\bibitem{leung} P.W. Leung and N.W. Lam, \prb {\bf 53}, 2213 (1996).
\bibitem{caprio1} L. Capriotti, D.J. Scalapino, and S.R. White \prl {\bf 93}, 
   177004 (2004).
\bibitem{poilblanc3} M. Mambrini, A. L\"auchli, D. Poilblanc, and F. Mila, 
   \prb {\bf 74}, 144422 (2006).
\bibitem{weber} C. Weber {\it et al.}, \prl {\bf 91}, 177202 (2003).
\bibitem{note} On the $8 \times 8$ lattice, different spin configurations,
   compatible with this conditions, have tiny energy differences among them.

\end{thebibliography}
\end{document}